\documentclass[conference]{IEEEtran}
\IEEEoverridecommandlockouts
\usepackage{cite}
\usepackage[T1]{fontenc}
\usepackage[utf8]{inputenc}
\usepackage{amsmath,amssymb,amsfonts}
\usepackage{graphicx}
\usepackage{textcomp}
\usepackage{xcolor}
\usepackage{booktabs}
\usepackage{multirow}
\usepackage{array}
\usepackage{float}
\usepackage{algorithm}
\usepackage{algpseudocode}
\usepackage{url}
\usepackage{placeins}
\usepackage{dblfloatfix}

\makeatletter
\renewcommand{\footnoterule}{%
  \kern -3pt
  \hrule \@width \columnwidth
  \kern 2.6pt}
\makeatother

\newcolumntype{L}[1]{>{\raggedright\arraybackslash}p{#1}}
\newcolumntype{C}[1]{>{\centering\arraybackslash}p{#1}}
\newcolumntype{R}[1]{>{\raggedleft\arraybackslash}p{#1}}

\providecommand{\tightlist}{%
  \setlength{\itemsep}{0pt}\setlength{\parskip}{0pt}}

\title{Few-Shot Open-Set Audio Classification via Transductive Prototype Refinement and Class Logit Enhancement}

\author{%
\IEEEauthorblockN{Tianyan Deng}
\IEEEauthorblockA{\textit{School of Mathematics}\\
\textit{South China University of Technology}\\
Guangzhou, China\\
202330400581@mail.scut.edu.cn} \\
\IEEEauthorblockN{Rui Gao}
\IEEEauthorblockA{\textit{School of Electronic and Information Engineering}\\
\textit{South China University of Technology}\\
Guangzhou, China\\
202330350401@mail.scut.edu.cn}
\and
\IEEEauthorblockN{Yanxiong Li\textsuperscript{*}\thanks{$^*$~Corresponding author: Yanxiong Li (eeyxli@scut.edu.cn).}}
\IEEEauthorblockA{\textit{School of Electronic and Information Engineering}\\
\textit{South China University of Technology}\\
Guangzhou, China\\
eeyxli@scut.edu.cn} \\
\IEEEauthorblockN{Jiahao Du}
\IEEEauthorblockA{\textit{School of Electronic and Information Engineering}\\
\textit{South China University of Technology}\\
Guangzhou, China\\
202330350311@mail.scut.edu.cn}
}

\IEEEaftertitletext{\vspace{-10pt}}

\algrenewcommand{\algorithmiccomment}[1]{\hfill // #1}

\begin{document}
\maketitle

\begin{abstract}
Few-shot Open-set audio classification requires classifying
query samples from known classes with a few labeled support samples while
rejecting query samples from unknown classes.
Transductive inference jointly observes the full unlabeled query set to
improve prototype estimation, yet standard transductive updates do not
distinguish known from unknown query samples, leaving prototypes vulnerable to
open-set contamination.
Drawing on latent-inlierness weighting and decoupled scoring for unknown-class samples,
we propose a two-phase transductive method operating over a frozen audio encoder.
First, each query sample is assigned a latent inlierness score that
down-weights likely unknown-class samples, so that prototype refinement
is driven primarily by known-class evidence.
The refined prototypes are then directly optimized on a transductive
loss combining support cross-entropy, inlierness-weighted conditional
entropy minimization, and inlierness-weighted marginal entropy
maximization, while
open-set rejection uses a prior-adaptive free-energy score that adjusts
its threshold with the prior proportion of unknown-class samples, decoupling detection
from classification.
Experiments on three audio datasets show our method achieves 
state-of-the-art results for few-shot open-set audio classification under multiple experimental conditions.
Code is available at \url{https://github.com/Gostyan/ROLE}
\end{abstract}


\section{Introduction}\label{introduction}

Real-world few-shot audio classification~\cite{ref41,ref42,ref43,ref44} rarely satisfies
the closed-set assumption: query samples routinely include sounds outside classes of support set.
In the transductive setting, a model observes the full unlabeled query set jointly
during inference, which improves episode-level prototype estimation~\cite{ref14,ref15}.
However, standard transductive updates treat all query samples equally and cannot
distinguish known-class from unknown-class samples, so prototype estimation degrade as the
unknown-class proportion grows~\cite{ref1,ref2,ref28}.
Existing transductive few-shot methods mainly target closed-set accuracy, while
dedicated open-set methods do not fully exploit query set structure.
This tension is especially pronounced in audio, where environmental sounds span
an open-ended category space and deployment conditions are highly variable.

Few-shot learning provides metric- and meta-learning foundations via
Prototypical Networks~\cite{ref11}, MAML~\cite{ref12}, and embedding
baselines~\cite{ref13}.
Few-shot open-set methods such as
OPP~\cite{ref34}, MET~\cite{ref35}, Glocal~\cite{ref39}, TANE~\cite{ref40},
and AISP~\cite{refAISP}
improve inductive rejection through various prototype and scoring strategies.
Transductive methods, such as TIM~\cite{ref14} and Prototype Rectification~\cite{ref15},
improve closed-set class estimation but are not designed for
open-set episodes. Contextual transductive FSOR~\cite{ref28} is the closest prior
transductive open-set extension.
OSLO~\cite{ref1} introduces latent-inlierness weighting for prototype estimation
and EOL~\cite{ref2} proposes decoupled outlier-logit scoring.
Few-shot open-set keyword-spotting methods tackle open-set challenges through
multi-stage training~\cite{ref36}, on-device calibration~\cite{ref37},
and cross-domain reprojection~\cite{ref38}, but remain task-specific and do not
address transductive prototype contamination for general audio classification.

We propose Refinement-based Outlier-Logit Enhancement (ROLE),
a transductive algorithm for Few-shot Open-set Audio Classification (FOAC)  operating on a frozen audio encoder.
Our contributions are as follows.
\begin{itemize}
\tightlist
\item We propose ROLE, a unified transductive inference procedure that
  couples latent-inlierness-guided prototype refinement with a
  prior-adaptive free-energy rejection score in a single episode-level
  inference rule, requiring only a frozen pre-trained encoder with no
  episodic meta-training or backbone fine-tuning.
\item Evaluated on ESC-50, FSD-Kaggle2018, and UrbanSound8K, under 5-way 1-shot/5-shot and
  20\%/50\%/80\% outlier ratios, our method achieves the highest AUROC scores in 10 out of 18
  evaluation settings, with macro-mean AUROC of 85.88/92.22 vs.\ 81.15/90.26
  for the strongest baseline MET.
\end{itemize}

\section{Method}\label{raft-method}

\begin{figure*}[t]
\centering
\includegraphics[width=\textwidth]{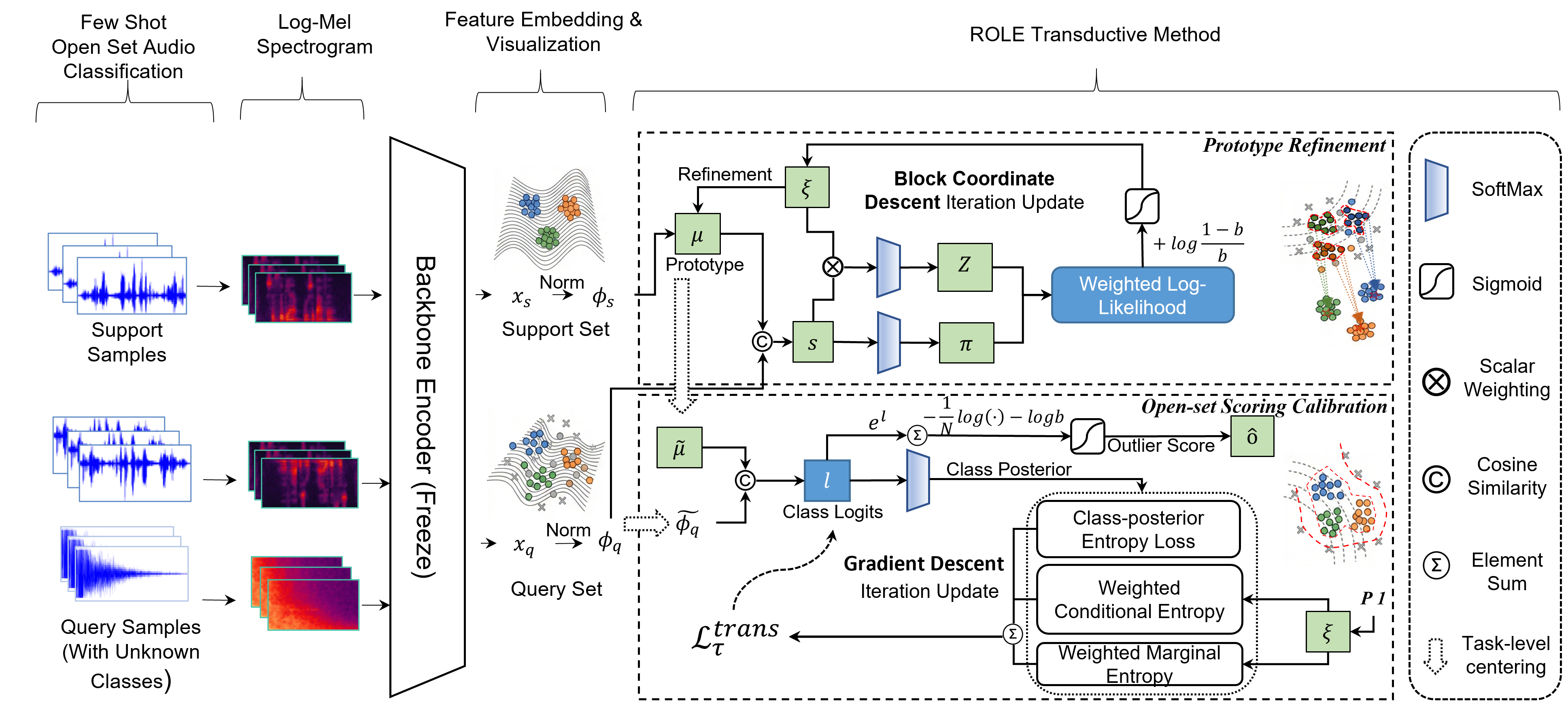}
\caption{Overview of ROLE. A frozen audio encoder maps support and query audio to normalized embeddings. Phase 1 performs inlierness-guided prototype refinement. Phase 2 centers episode geometry, optimizes prototypes with $\xi$-weighted transductive losses, and uses prior-adaptive free-energy rejection decoupled from classification.}
\label{fig:role-overview}
\end{figure*}

\subsection{Problem Formulation}\label{problem-formulation}

We consider a transductive $N$-way, $K$-shot open-set episode.
The support set $\mathcal{S}=\{(x_{s_i}, y_{s_i})\}_{i=1}^{N_s}$
($N_s=NK$) contains labeled examples from $N$ novel classes, and the
unlabeled query set $\mathcal{Q}=\{x_{q_j}\}_{j=1}^{N_q}$ may contain
both known-class samples (\emph{inliers}) and unknown-class samples
(\emph{outliers}). The full query set is observed jointly during
inference. The goal is to classify inlier query samples into one of $N$
support classes and reject outliers with a scalar outlier score.
We write $[N]=\{1,\ldots,N\}$ for the inlier label set.

ROLE operates on a frozen pre-trained audio encoder $f(\cdot)$ that maps
each waveform to a normalized embedding $\phi=f(x)/\|f(x)\|$. We write
$\phi_{s_i}$ and $\phi_{q_j}$ for support and query embeddings
respectively. ROLE optimizes only episode-specific parameters: class
prototypes $\{\mu_k\}_{k=1}^N$, re-initialized per episode.

\subsection{Design Principles}\label{design-principles}

ROLE is structured as a two-phase pipeline, and each phase addresses a
distinct challenge in transductive open-set classification.

\textbf{Prototype bias from undifferentiated transduction.}
Standard transductive updates (e.g., TIM~\cite{ref14}, Prototype
Rectification~\cite{ref15}) incorporate all query evidence equally into
prototype estimation.
In open-set episodes, unknown-class query samples inevitably participate in these
updates, pulling prototypes away from the true known-class centers and the
bias grows with the unknown-class ratio.
Phase~1 mitigates this by introducing a per-query latent inlierness score:
query samples assessed as unlikely to belong to any known class are automatically
down-weighted, so that primarily known-class evidence refines the prototypes.

\textbf{Conflated classification and rejection under a single softmax.}
A natural approach uses the softmax posterior over known classes for both
classification and open-set rejection.
However, softmax normalizes only over known classes and cannot express
``this query sample belongs to none of them'': an unknown-class query sample equidistant
from all prototypes receives a uniform but non-negligible posterior,
appearing similar to a genuinely ambiguous inlier.
Phase~2 decouples the two objectives by deriving classification from a
class-wise softmax and open-set rejection from a sigmoid on the negative
log-mean-exp of the same logits, which explicitly measures
whether total class evidence justifies considering the query sample as known-class.

Figure~\ref{fig:role-overview} summarizes the resulting pipeline.

\subsection{Unified Transductive Inference}\label{unified-episode-adaptation-rule}

Given a labeled support set and an unlabeled open-set query set, ROLE first
refines prototypes under latent inlierness and then further optimizes them
on a transductive loss with a prior-adaptive decoupled scoring objective.

\textbf{Phase 1: Inlierness-guided prototype refinement.}
Phase~1 refines class prototypes using query-batch evidence while
preventing unknown-class query samples from biasing the result.
We write $\operatorname{Norm}(\cdot)=(\cdot)/\|\cdot\|$ for $\ell_2$
normalization.
Prototypes are initialized from support means:
$\mu_k^{(0)}=\operatorname{Norm}(\sum_{i:y_{s_i}=k}\phi_{s_i})$.
Each query sample $j$ carries a latent inlierness score $\xi_j\in(0,1)$
(initialized to $\xi_j^{(0)}=1$), representing the model's belief that
query sample $j$ belongs to some known class.
Following the block-coordinate descent of OSLO~\cite{ref1}, three
variables---soft assignments $Z$, inlierness $\xi$, and prototypes
$\mu$---are updated alternately for $T_{\mathrm{boot}}$ iterations with
support assignments held fixed.
At iteration $t$, let
$s_{jk}^{(t)}=\tau\langle\mu_k^{(t)},\phi_{q_j}\rangle$ denote the
temperature-scaled cosine similarity between query sample $j$ and prototype $k$, and
$\pi_{jk}^{(t)}=\operatorname{Softmax}_{k\in[N]}(s_{jk}^{(t)})$ the
posterior probability that query sample $j$ belongs to class $k$.

Soft assignments are modulated by inlierness rather than committed to
hard pseudo-labels, which would propagate assignment errors directly
into prototypes:
\begin{equation}
Z_{jk}^{(t+1)} = \operatorname{Softmax}_{k \in [N]}\!\left(
\xi_j^{(t)}\,s_{jk}^{(t)}
\right)
\end{equation}
Here $\xi_j$ acts as an inverse temperature: when $\xi_j\!\approx\!1$
the assignment peaks at the nearest prototype, when
$\xi_j\!\approx\!0$ it flattens to uniform, neutralizing the query sample's
contribution.
The inlierness score is then updated via the expected log-posterior
under the soft assignment, measuring how well the known-class structure
explains query sample $j$:
\begin{equation}
\xi_j^{(t+1)} = \operatorname{Sigmoid}\!\left(
\frac{1}{\lambda_{\xi}} \sum_k Z_{jk}^{(t+1)}
\log \pi_{jk}^{(t)} + \log\frac{1-b}{b}
\right)
\end{equation}
where $\lambda_{\xi}>0$ controls sensitivity and $b$ is the prior outlier
proportion of the episode.
The bias term $\log\tfrac{1-b}{b}$ shifts $\xi_j$ downward when $b$ is large:
at high outlier ratios, the default $\sigma(0)=0.5$ would place many
outliers at an ambiguous midpoint, making them hard to suppress,
the prior shift pushes the threshold so that only query samples with strong
known-class evidence retain high inlierness.
Prototypes are re-estimated by combining fixed support geometry with
inlierness-filtered query evidence:
\begin{equation}
\mu_k^{(t+1)} = \operatorname{Norm}\!\left(
\sum_{i:\,y_{s_i}=k} \phi_{s_i}
+ \sum_{j=1}^{N_q} \xi_j^{(t+1)}\,Z_{jk}^{(t+1)}\,\phi_{q_j}
\right)
\end{equation}
The product $\xi_j Z_{jk}$ provides double gating: only query samples that
are both likely inliers and confidently assigned to class $k$ influence
$\mu_k$, preventing unknown-class query samples from shifting prototypes.

\textbf{Phase 2: Transductive prototype optimization with decoupled scoring.}
Phase~2 directly optimizes the class prototypes $\{\mu_k\}$ on the
episode-level transductive loss, further adapting the refined geometry to
the observed query distribution.
Because frozen encoders carry a shared bias from pre-training, we first
subtract the episode mean
$\bar{\phi}=\tfrac{1}{N_s+N_q}(\sum_i\phi_{s_i}+\sum_j\phi_{q_j})$,
yielding centered prototypes
$\tilde{\mu}_k=\mu_k^{(T_{\mathrm{boot}})}-\bar{\phi}$ and centered query samples
$\tilde{\phi}_{q_j}=\phi_{q_j}-\bar{\phi}$.
The logit for query sample $j$ on class $k$ is
$l_{jk}=\tau\langle\tilde{\phi}_{q_j},\tilde{\mu}_k\rangle$,
the temperature-scaled cosine similarity between the centered query sample and prototype,
where $\tau>0$ is a fixed temperature shared across both phases.

Following EOL~\cite{ref2}, classification and rejection are decoupled
rather than sharing a single softmax:
\begin{equation}
\begin{aligned}
\hat{o}_j
&= \operatorname{Sigmoid}\!\left(-\log\tfrac{1}{N}\sum_{k=1}^{N} e^{l_{jk}} - \log b\right), \\
p_{jk} &= \operatorname{Softmax}_{k \in [N]}(l_{jk}).
\end{aligned}
\end{equation}
The class posterior $p_{jk}$ gives the predicted class
$\hat{y}_j=\arg\max_k\,l_{jk}$.
The outlier score $\hat{o}_j$ is derived from the negative log-mean-exp of
the logits: when no class logit is high, the sum $\sum_k e^{l_{jk}}$ is small
relative to $N$, so $-\log\frac{1}{N}\sum_k e^{l_{jk}}$ is large and $\hat{o}_j$
is correspondingly large.
The prior-adaptive shift $-\log b$ further lowers the free-energy threshold
required for inlier acceptance as the outlier proportion $b$ grows, so that the
boundary between inlier and outlier decisions adapts automatically to episode difficulty.

The prototypes $\{\tilde{\mu}_k\}$ are optimized over $T_{\mathrm{cal}}$ gradient
steps on a transductive loss:
\begin{equation}
\begin{aligned}
\mathcal{L}^{\mathrm{trans}} = {}&
-\frac{1}{N_s}\sum_{i=1}^{N_s}\log p_{i,y_{s_i}}
+ \lambda_q \frac{\sum_j \xi_j H(p_j)}{\sum_j \xi_j} \\
&+ \lambda_{\mathrm{ma}} \sum_k \hat{p}_k \log \hat{p}_k
\end{aligned}
\end{equation}
where $\hat{p}_k = \bigl(\sum_j \xi_j p_{jk}\bigr)\big/\bigl(\sum_j \xi_j\bigr)$
is the $\xi$-weighted marginal class distribution; $\lambda_q$ and $\lambda_{\mathrm{ma}}>0$
are hyperparameters balancing the three objectives.
The three terms are as follows.
The support cross-entropy anchors prototype updates to labeled data.
The $\xi_j$-weighted conditional entropy minimization encourages sharp per-query
predictions for likely inliers while ignoring outliers ($\xi_j\!\approx\!0$),
where the per-query entropy is
\begin{equation}
H(p_j) = -\sum_k p_{jk}\log p_{jk}.
\end{equation}
The $\xi$-weighted marginal entropy maximization: since $\hat{p}_k\log\hat{p}_k\le 0$,
minimizing $+\lambda_{\mathrm{ma}}\sum_k\hat{p}_k\log\hat{p}_k$ pushes $\hat{p}$
toward the uniform distribution, encouraging the prototype geometry to assign
inlier query samples evenly across known classes and preventing collapse to a dominant class.

Unlike OSLO/EOL, ROLE uses inlierness-gated TIM-style optimization before
prior-adaptive free-energy scoring.

Algorithm~\ref{alg:role} summarizes the complete procedure of ROLE.

\begin{algorithm}[t]
\caption{ROLE}\label{alg:role}
\begin{algorithmic}[1]
\Require Support set $\mathcal{S}$, query set $\mathcal{Q}$, frozen encoder $f$
\Ensure Predicted label $\hat{y}_j$ and outlier score $\hat{o}_j$ for each query sample $j$
\State Encode and $\ell_2$-normalise all embeddings; init $\mu_k^{(0)}$ from support means, $\xi_j^{(0)}{=}1$
\For{$t=0,\ldots,T_{\mathrm{boot}}-1$} \Comment{Phase 1}
    \State $s_{jk}^{(t)}\!=\!\tau\langle\mu_k^{(t)},\phi_{q_j}\rangle$,\quad $\pi_{jk}^{(t)}\!=\!\operatorname{Softmax}_k(s_{jk}^{(t)})$
    \State $Z_{jk}^{(t+1)}\!\leftarrow\!\operatorname{Softmax}_k\!\bigl(\xi_j^{(t)}\,s_{jk}^{(t)}\bigr)$
    \State $\xi_j^{(t+1)}\!\leftarrow\!\sigma\!\Bigl(\tfrac{1}{\lambda_\xi}\sum_k Z_{jk}^{(t+1)}\log\pi_{jk}^{(t)}+\log\tfrac{1-b}{b}\Bigr)$
    \State $\mu_k^{(t+1)}\!\leftarrow\!\operatorname{Norm}\!\bigl(\textstyle\sum_{i:y_{s_i}=k}\phi_{s_i}+\sum_j\xi_j^{(t+1)}Z_{jk}^{(t+1)}\phi_{q_j}\bigr)$
\EndFor
\State Centre all embeddings: $\tilde{\phi}\leftarrow\phi-\bar{\phi}$ \Comment{Phase 2}
\For{$t=0,\ldots,T_{\mathrm{cal}}-1$}
    \State $l_{jk}\!=\!\tau\langle\tilde{\phi}_{q_j},\tilde{\mu}_k\rangle$;\quad minimise $\mathcal{L}^{\mathrm{trans}}$ w.r.t.\ $\{\tilde{\mu}_k\}$
\EndFor\Statex\Return $\hat{y}_j\!=\!\arg\max_k\,l_{jk}$,\quad $\hat{o}_j\!=\!\sigma\!\bigl(-\log\tfrac{1}{N}\sum_k e^{l_{jk}}-\log b\bigr)$
\end{algorithmic}
\end{algorithm}

\section{Experiments}\label{experiments}

\subsection{Datasets}\label{datasets}

We evaluate different methods on three audio datasets: ESC-50~\cite{ref22}, FSD-Kaggle2018~\cite{ref24} and UrbanSound8K~\cite{ref23}.
The encoder is pre-trained on AudioSet~\cite{ref21} and kept frozen.
No dataset contributes a training split. All clips are used for episodic evaluation.
Each dataset is divided into 5 folds. Episodes are drawn from
the 4 non-held-out folds with the held-out fold rotated across iterations.

\textbf{ESC-50}~\cite{ref22} contains 2,000 five-second environmental sound
clips across 50 classes (40 clips/class) with 5 built-in cross-validation folds.
We designate 25 classes (indices 0--24) as the inlier pool and the remaining
25 classes (indices 25--49) as the outlier pool.
Each episode samples 5 inlier and 5 outlier classes from their respective pools.

\textbf{FSD-Kaggle2018}~\cite{ref24} contains 9,473 variable-length clips
across 41 general-purpose sound-event classes (94--300 clips/class).
Because the dataset has no built-in folds, we assign 5 stratified folds
deterministically (seed\,=\,0).
We use 20 classes (indices 0--19) as inlier pool and 21 classes (indices
20--40) as outlier pool, each episode samples 5\,+\,5 from these pools.

\textbf{UrbanSound8K}~\cite{ref23} contains 8,732 urban sound clips across 10 classes (374--1,000 clips/class).
The original 10 folds are remapped to 5 by merging consecutive pairs
(folds 1--2\,$\to$\,1, 3--4\,$\to$\,2, \ldots).
With 5 inlier classes (indices 0--4) and 5 outlier classes (indices 5--9),
every episode uses all 10 classes.

\subsection{Experimental Setup}\label{experimental-setup}

Each configuration uses 5-way open-set
episodes with 5 inlier classes and 5 outlier classes drawn from their
respective pools. We evaluate both 1-shot and 5-shot support samples.
Support samples come exclusively from inlier classes. All query samples
are unlabeled at test time. Each inlier class contributes 4 query
samples, whereas each outlier class contributes 1, 4, or 16 query
samples, yielding 20\%, 50\%, and 80\% outlier ratios.
Under the 1-shot setting, this gives 5 support samples per episode. Under
5-shot, 25 support samples per episode. Query set sizes are 25, 40, and 100 for the three outlier
ratios, regardless of shot count.

We follow a 5-fold cross-fold protocol with the held-out fold excluded
from episode sampling. 300 episodes are drawn per fold, yielding 1,500
episodes per configuration. Because ROLE is a transductive inference
method without cross-episode training, support and query samples
come from the non-held-out folds of the same dataset. Across the full
benchmark (8 baselines $+$ ROLE, 3 datasets, 2 shot settings, 3 outlier
ratios), this amounts to $9 \times 3 \times 2 \times 3 \times 1{,}500
= 243{,}000$ episode-level inferences.

We report three metrics~\cite{ref33}:
inlier classification accuracy (Acc), measuring the fraction of
correctly classified known-class query samples,
area under the receiver operating characteristic curve (AUROC),
which evaluates the ranking quality of the outlier score across all
operating thresholds, and area under the precision--recall curve (AUPR),
which emphasizes detection performance when the positive (outlier) class
is rare or dominant depending on the outlier ratio.
Table~\ref{tab:detailed} shows per-dataset results across all outlier ratios and shot
settings. Table~\ref{tab:means} presents the macro-average over all three datasets,
further averaged over the three outlier ratios, under 1-shot and 5-shot protocols.
Both tables arrange methods by row.

All methods share the same pre-trained Audio
Spectrogram Transformer (AST \cite{ref7}, pre-trained on AudioSet
\cite{ref21}). Following standard model-agnostic transductive FSL
practice~\cite{ref14,ref15}, the encoder is kept fixed during
inference so that the comparison isolates differences in the FOAC
algorithm itself.

We compare ROLE against four transductive baselines,
OSTIM \cite{ref33}, OSLO \cite{ref1},
EOL \cite{ref2}, and OPP-T \cite{ref34}, as well as five inductive
baselines, OPP-I \cite{ref34}, MET \cite{ref35}, Glocal \cite{ref39},
TANE \cite{ref40}, and AISP \cite{refAISP}. All methods use the same frozen AST embeddings
and identical episodic construction so that differences can be
attributed to the open-set inference rule rather than to backbone
changes.
For OPP-I/OPP-T, since the AST backbone cannot be retrained, base-class weight vectors are computed
as the mean of frozen AST embeddings over all training-fold samples per class, omitting the
backbone fine-tuning stage of the original paper.
For MET, Glocal, and TANE, all three share the same cosine ProtoNet classifier. Only the
outlier scoring function differs, and no per-method temperature tuning is applied, which
accounts for their identical Acc values.
AISP is evaluated under a class-split protocol (60\% base-class meta-training / 40\% novel-class
evaluation). UrbanSound8K is excluded for AISP as its 10-class pool yields only 4 novel classes
after splitting, which is insufficient for 5-way episodes.

\begin{table*}[!t]
\centering
\caption{Detailed per-dataset results. Metrics: Acc/AUROC/AUPR (\%).}
\label{tab:detailed}
\tiny
\setlength{\tabcolsep}{1.2pt}
\renewcommand{\arraystretch}{0.96}
\resizebox{\textwidth}{!}{%
\begin{tabular}{@{}ll*{6}{c}|*{6}{c}|*{6}{c}@{}}
\toprule\noalign{}
\multirow{3}{*}{Method} & \multirow{3}{*}{\shortstack{Outlier\\Ratio}} & \multicolumn{6}{c}{ESC-50} & \multicolumn{6}{c}{FSD-Kaggle2018} & \multicolumn{6}{c}{UrbanSound8K} \\
\cmidrule(lr){3-8}\cmidrule(lr){9-14}\cmidrule(lr){15-20}
& & \multicolumn{3}{c}{5-way 1-shot} & \multicolumn{3}{c}{5-way 5-shot} & \multicolumn{3}{c}{5-way 1-shot} & \multicolumn{3}{c}{5-way 5-shot} & \multicolumn{3}{c}{5-way 1-shot} & \multicolumn{3}{c}{5-way 5-shot} \\
\cmidrule(lr){3-5}\cmidrule(lr){6-8}\cmidrule(lr){9-11}\cmidrule(lr){12-14}\cmidrule(lr){15-17}\cmidrule(lr){18-20}
& & Acc & AUROC & AUPR & Acc & AUROC & AUPR & Acc & AUROC & AUPR & Acc & AUROC & AUPR & Acc & AUROC & AUPR & Acc & AUROC & AUPR \\
\midrule\noalign{}
OSTIM~\cite{ref33} & \multirow{10}{*}{20\%} & \textbf{98.41} & 94.44 & 82.81 & \textbf{99.46} & 96.20 & 87.65 & \textbf{88.05} & 80.48 & 54.51 & \textbf{94.26} & 84.62 & 60.60 & \textbf{80.13} & 67.30 & 39.03 & \textbf{90.84} & 74.73 & 45.77 \\\
OSLO~\cite{ref1} &  & 97.54 & \textbf{97.98} & \textbf{94.17} & 99.27 & 98.53 & 96.15 & 83.91 & \textbf{86.32} & \textbf{67.86} & 92.30 & 89.95 & 74.04 & 72.43 & \textbf{73.99} & \textbf{50.07} & 87.78 & 77.96 & 55.89 \\
EOL~\cite{ref2} &  & 96.10 & 71.52 & 38.11 & 98.44 & 88.57 & 62.65 & 85.55 & 65.13 & 36.29 & 93.05 & 79.19 & 48.82 & 76.23 & 57.90 & 31.62 & 89.04 & 69.08 & 39.98 \\
OPP-I~\cite{ref34} &  & 96.32 & 88.68 & 73.24 & 99.20 & 92.01 & 80.74 & 81.39 & 71.05 & 46.01 & 92.10 & 73.88 & 48.96 & 69.28 & 61.42 & 37.47 & 87.28 & 61.86 & 38.24 \\
OPP-T~\cite{ref34} &  & 98.23 & 88.49 & 73.64 & 99.30 & 90.80 & 78.47 & 79.68 & 70.60 & 46.19 & 92.52 & 72.82 & 48.11 & 75.99 & 61.62 & 37.68 & 88.08 & 61.48 & 38.02 \\
MET~\cite{ref35} &  & 96.32 & 94.49 & 85.85 & 99.20 & 98.27 & 95.43 & 81.39 & 80.28 & 58.04 & 92.10 & 90.39 & 75.86 & 69.28 & 68.53 & 44.53 & 87.28 & 81.79 & \textbf{60.13} \\
Glocal~\cite{ref39} &  & 96.32 & 94.66 & 85.42 & 99.20 & 98.22 & 95.09 & 81.39 & 78.68 & 54.80 & 92.10 & 87.75 & 68.91 & 69.28 & 67.51 & 43.18 & 87.28 & 74.00 & 50.33 \\
TANE~\cite{ref40} &  & 96.32 & 93.51 & 82.44 & 99.20 & 97.47 & 93.06 & 81.39 & 76.18 & 51.51 & 92.10 & 84.13 & 62.72 & 69.28 & 65.19 & 40.98 & 87.28 & 69.02 & 45.55 \\
AISP~\cite{refAISP} &  & 93.74 & 87.14 & 75.34 & 96.21 & 94.32 & 86.90 & 69.21 & 75.66 & 56.43 & 84.59 & 81.44 & 61.61 & \multicolumn{6}{c}{---} \\
ROLE (Ours) &  & 97.94 & 96.09 & 88.99 & 99.34 & \textbf{98.69} & \textbf{96.45} & 85.79 & 82.55 & 60.69 & 92.91 & \textbf{91.74} & \textbf{78.56} & 75.41 & 70.12 & 43.87 & 88.46 & \textbf{82.08} & 59.85 \\
\midrule\noalign{}
OSTIM~\cite{ref33} & \multirow{10}{*}{50\%} & \textbf{96.97} & 70.03 & 70.57 & 99.19 & 76.32 & 75.61 & \textbf{84.63} & 71.84 & 71.35 & 92.66 & 76.75 & 75.83 & \textbf{75.54} & 53.00 & 57.77 & 88.38 & 60.81 & 63.26 \\
OSLO~\cite{ref1} &  & 96.87 & 76.74 & 74.87 & 99.13 & 92.40 & 91.09 & 83.31 & 73.80 & 72.65 & 92.28 & 82.82 & 80.93 & 70.42 & 54.05 & 55.33 & 86.75 & 66.54 & 65.04 \\
EOL~\cite{ref2} &  & 96.86 & 96.26 & 95.79 & 99.23 & \textbf{99.24} & 99.21 & 83.96 & 83.88 & 82.21 & 92.74 & \textbf{93.69} & \textbf{93.13} & 74.30 & 74.14 & 74.00 & 88.30 & \textbf{86.83} & \textbf{86.57} \\
OPP-I~\cite{ref34} &  & 96.07 & 89.47 & 88.73 & 99.27 & 91.82 & 91.37 & 81.34 & 71.44 & 70.30 & 92.55 & 74.35 & 73.28 & 68.59 & 61.22 & 62.50 & 87.31 & 61.99 & 63.76 \\
OPP-T~\cite{ref34} &  & 96.87 & 67.77 & 67.38 & 98.91 & 79.01 & 77.59 & 83.95 & 60.68 & 61.47 & 91.86 & 66.83 & 66.64 & 70.63 & 54.59 & 56.05 & 85.79 & 57.33 & 59.18 \\
MET~\cite{ref35} &  & 96.07 & 94.75 & 94.39 & 99.27 & 98.50 & 98.54 & 81.34 & 80.64 & 79.24 & 92.55 & 90.75 & 90.04 & 68.59 & 68.16 & 68.47 & 87.31 & 82.16 & 81.07 \\
Glocal~\cite{ref39} &  & 96.07 & 95.06 & 94.55 & 99.27 & 98.44 & 98.41 & 81.34 & 79.19 & 77.32 & 92.55 & 88.17 & 86.57 & 68.59 & 67.14 & 67.39 & 87.31 & 74.38 & 74.21 \\
TANE~\cite{ref40} &  & 96.07 & 93.93 & 93.22 & 99.27 & 97.73 & 97.60 & 81.34 & 76.61 & 74.80 & 92.55 & 84.57 & 82.89 & 68.59 & 64.82 & 65.51 & 87.31 & 69.34 & 70.21 \\
AISP~\cite{refAISP} &  & 90.13 & 87.10 & 86.69 & 98.42 & 94.73 & 94.93 & 80.13 & 71.81 & 71.66 & 88.73 & 81.23 & 80.69 & \multicolumn{6}{c}{---} \\
ROLE (Ours) &  & 96.70 & \textbf{97.22} & \textbf{97.31} & \textbf{99.32} & 99.15 & \textbf{99.22} & 83.73 & \textbf{85.74} & \textbf{85.26} & \textbf{93.05} & 93.05 & 92.77 & 73.73 & \textbf{74.90} & \textbf{75.02} & \textbf{88.39} & 84.85 & 84.54 \\
\midrule\noalign{}
OSTIM~\cite{ref33} & \multirow{10}{*}{80\%} & 96.33 & 34.37 & 74.11 & 98.78 & 43.80 & 78.14 & 81.25 & 57.05 & 84.56 & 91.58 & 62.94 & 86.84 & 71.90 & 35.80 & 75.88 & 85.86 & 42.53 & 85.68 \\
OSLO~\cite{ref1} &  & 82.26 & 31.62 & 66.30 & 95.04 & 37.58 & 72.95 & 73.87 & 40.07 & 76.18 & 87.77 & 55.37 & 82.63 & 56.09 & 28.72 & 70.04 & 72.77 & 35.63 & 72.66 \\
EOL~\cite{ref2} &  & 96.49 & 96.42 & 98.99 & 98.98 & 98.92 & 99.71 & 82.24 & 87.32 & 96.02 & 92.27 & \textbf{93.95} & \textbf{98.22} & 73.74 & 77.06 & 92.58 & 87.45 & \textbf{87.52} & \textbf{96.27} \\
OPP-I~\cite{ref34} &  & 96.19 & 88.41 & 95.97 & \textbf{99.05} & 91.84 & 97.30 & 80.93 & 71.70 & 89.05 & 92.33 & 73.43 & 89.99 & 68.96 & 61.64 & 85.78 & 87.05 & 61.91 & 86.09 \\
OPP-T~\cite{ref34} &  & 71.48 & 30.72 & 69.54 & 92.93 & 43.94 & 76.97 & 69.09 & 41.66 & 76.51 & 85.80 & 51.40 & 80.73 & 52.66 & 44.79 & 76.84 & 70.47 & 49.29 & 79.19 \\
MET~\cite{ref35} &  & 96.19 & 94.47 & 98.12 & \textbf{99.05} & 98.23 & 99.48 & 80.93 & 80.51 & 92.81 & 92.33 & 90.28 & 96.78 & 68.96 & 68.56 & 88.53 & 87.05 & 82.00 & 93.75 \\
Glocal~\cite{ref39} &  & 96.19 & 94.66 & 98.12 & \textbf{99.05} & 98.20 & 99.45 & 80.93 & 79.17 & 92.03 & 92.33 & 87.45 & 95.42 & 68.96 & 67.54 & 88.14 & 87.05 & 74.26 & 90.92 \\
TANE~\cite{ref40} &  & 96.19 & 93.42 & 97.65 & \textbf{99.05} & 97.43 & 99.18 & 80.93 & 76.66 & 90.96 & 92.33 & 83.81 & 94.01 & 68.96 & 65.29 & 87.28 & 87.05 & 69.24 & 89.16 \\
AISP~\cite{refAISP} &  & 89.58 & 88.26 & 96.14 & 96.26 & 95.31 & 98.64 & 63.66 & 72.87 & 90.13 & 87.06 & 82.10 & 93.87 & \multicolumn{6}{c}{---} \\
ROLE (Ours) &  & \textbf{96.56} & \textbf{98.00} & \textbf{99.45} & 99.00 & \textbf{99.25} & \textbf{99.80} & \textbf{82.97} & \textbf{88.88} & \textbf{96.60} & \textbf{92.86} & 93.83 & 98.19 & \textbf{73.86} & \textbf{79.38} & \textbf{93.45} & \textbf{88.07} & 87.33 & 96.15 \\
\bottomrule\noalign{}
\end{tabular}
}
\end{table*}

\subsection{Main Results}\label{main-results}

Tables~\ref{tab:detailed} and~\ref{tab:means} summarize the benchmark.
ROLE achieves the best AUROC in 10 out of the 18 dataset-specific settings.
At 20\% outlier ratios under 1-shot, OSLO leads in all three datasets:
with only a small number of query samples of unknown classes, prototype refinement alone
suffices and the benefit of decoupled scoring diminishes.
Table~\ref{tab:means} confirms ROLE's macro-average advantage:
ROLE attains 85.88\% (1-shot) and 92.22\% (5-shot) AUROC,
improving over MET by 4.73\% and 1.96\% respectively,
and leads inlier Acc under 5-shot (93.49\%).
Under 1-shot, OSTIM leads Acc narrowly (85.91\% vs.\ 85.19\%) but
trails ROLE by 23.18\% in AUROC, reflecting the gap between
closed-set transduction and open-set discriminability.
AISP, despite episodic meta-training, trails the training-free ROLE by 5.41\% in macro AUROC.

The per-method trends reveal complementary failure modes that ROLE
avoids. OSLO refines prototypes effectively at low contamination but
lacks a decoupled scoring head, so its AUROC collapses under 80\%
outliers (e.g.\ 31.62\% on ESC-50 1-shot) where corrupted prototypes
directly degrade the coupled softmax. Conversely, EOL applies decoupled
scoring without prototype refinement; it excels at high outlier ratios
yet underperforms at 20\% (e.g.\ 71.52\% AUROC on ESC-50 1-shot) because
its support-only prototypes are too coarse for reliable logit
calibration. The inductive methods MET~\cite{ref35}, Glocal~\cite{ref39}, and
TANE~\cite{ref40} adopt distinct rejection criteria but share a
structural limitation: their rejection boundaries are calibrated solely
from support examples and remain fixed at test time, causing
degradation as the episode outlier ratio shifts.
ROLE sidesteps these failure modes by jointly refining prototypes with
inlierness gating and adapting rejection to the episode outlier prior,
although acoustically similar unknown sounds may still influence refinement.

\begin{table}[!t]
\centering
\caption{Macro-averaged results over ESC-50, FSD-Kaggle2018, and UrbanSound8K,
further averaged over three outlier ratios per dataset. Acc/AUROC/AUPR (\%).}
\label{tab:means}
\small
\setlength{\tabcolsep}{3pt}
\renewcommand{\arraystretch}{0.98}
\begin{tabular}{@{}lcccccc@{}}
\toprule\noalign{}
\multirow{2}{*}{Method} & \multicolumn{3}{c}{5-way 1-shot} & \multicolumn{3}{c}{5-way 5-shot} \\
\cmidrule(lr){2-4}\cmidrule(lr){5-7}
& Acc & AUROC & AUPR & Acc & AUROC & AUPR \\
\midrule\noalign{}
OSTIM~\cite{ref33}  & \textbf{85.91} & 62.70 & 67.84 & 93.45 & 68.74 & 73.26 \\
OSLO~\cite{ref1}    & 79.63 & 62.59 & 69.72 & 90.34 & 70.75 & 76.82 \\
EOL~\cite{ref2}     & 85.05 & 78.85 & 71.73 & 93.28 & 88.55 & 80.51 \\
OPP-I~\cite{ref34}  & 82.12 & 73.89 & 72.12 & 92.90 & 75.90 & 74.41 \\
OPP-T~\cite{ref34}  & 77.62 & 57.88 & 62.81 & 89.52 & 63.66 & 67.21 \\
MET~\cite{ref35}    & 82.12 & 81.15 & 78.89 & 92.90 & 90.26 & 87.90 \\
Glocal~\cite{ref39} & 82.12 & 80.40 & 77.88 & 92.90 & 86.76 & 84.37 \\
TANE~\cite{ref40}   & 82.12 & 78.40 & 76.04 & 92.90 & 83.64 & 81.60 \\
AISP~\cite{refAISP} & 81.07 & 80.47 & 79.40 & 91.88 & 88.19 & 86.11 \\
ROLE (Ours)        & 85.19 & \textbf{85.88} & \textbf{82.29} & \textbf{93.49} & \textbf{92.22} & \textbf{89.50} \\
\bottomrule\noalign{}
\end{tabular}
\end{table}

\FloatBarrier
\subsection{Ablation Study}\label{ablation-study}

Table~\ref{tab:ablation} evaluates four component variants of ROLE on
ESC-50 and FSD-Kaggle2018, macro-averaged across three outlier ratios
(20\%, 50\%, 80\%) and both shot settings.
UrbanSound8K is omitted because its 10 classes make each episode cover all inlier/outlier classes, limiting ablation representativeness.

\begin{table}[t]
\centering
\caption{Ablation results macro-averaged over ESC-50 and FSD-Kaggle2018,
three outlier ratios, and both shot settings.
Format: mean\,$\pm$\,95\,\% CI (\%).}
\label{tab:ablation}
\small
\setlength{\tabcolsep}{5pt}
\renewcommand{\arraystretch}{1.0}
\begin{tabular}{@{}lcc@{}}
\toprule\noalign{}
Variant & AUROC & Acc \\
\midrule\noalign{}
ROLE                                 & \textbf{93.82}\,$\pm$\,\textbf{0.16} & \textbf{93.53}\,$\pm$\,\textbf{0.29} \\
w/o Phase~1                          & 91.91\,$\pm$\,0.16          & 91.88\,$\pm$\,0.28 \\
w/o $\xi$-gate ($\xi_j\!\equiv\!1$) & 71.77\,$\pm$\,0.29          & 90.01\,$\pm$\,0.45 \\
w/o Phase~2                          & 92.53\,$\pm$\,0.17          & 92.81\,$\pm$\,0.32 \\
w/o $b$-prior                        & 89.83\,$\pm$\,0.15          & 90.09\,$\pm$\,0.30 \\
\bottomrule\noalign{}
\end{tabular}
\end{table}

The ablation reveals a clear hierarchy of contributions.
Removing $\xi$-gating causes by far the largest AUROC drop
($-$22.05\%), confirming that inlierness gating is the most critical
mechanism: without it, Phase~1 propagates outlier contributions
uniformly into prototypes, severely corrupting the logit geometry.
Removing the prior-adaptive shift (w/o $b$-prior) causes the
next-largest degradation ($-$3.99\% AUROC, $-$3.44\% Acc), showing
that calibrating the inlierness threshold to the episode outlier
proportion is essential for both detection and classification.
Removing Phase~1 entirely yields moderate drops ($-$1.91\% AUROC,
$-$1.65\% Acc), indicating that Phase~2 can partially recover from
a coarser initialisation.
Disabling Phase~2 causes the smallest AUROC drop ($-$1.29\%) but
reduces Acc by 0.72\%, confirming that the transductive
optimisation further corrects few-shot class geometry beyond what
Phase~1 alone achieves.

\FloatBarrier
\section{Conclusions}\label{conclusions}

We presented ROLE, a transductive inference procedure for few-shot
open-set audio classification that couples inlierness-guided prototype
refinement with a prior-adaptive free-energy rejection score and
a transductive prototype optimization objective.
This design directly addresses the core tension of transductive FOAC:
query set transduction benefits prototype estimation only when
query samples of unknown classes are identified and suppressed first.
Experiments show macro-averaged AUROC/AUPR gains, especially at
moderate-to-high outlier ratios; ablations verify that both stages are necessary.
All experiments use a fixed pre-trained AST encoder, and stronger encoder
adaptation and inlierness estimation remain directions for future work.

\section*{Acknowledgment}

This work was partly supported by the national natural science foundation of China (62371195, 62111530145), the exchange project of the 10th Meeting of China-Croatia Science and Technology Cooperation Committee (10-34), and the national undergraduate training program for innovation and entrepreneurship (202510561025 and 202510561026).
\bibliographystyle{IEEEtran}
\bibliography{references}

\end{document}